# Obtaining Nanocarbon from Local Raw Materials and Studying Its Textural and Sorption Properties

Hilola N. Xolmirzayeva[1], Normurot I. Fayzullayev[2]

[1]*Laboratory assistant, Department of Analytical Chemistry, Samarkand State University, Samarkand, Uzbekistan*
*ORCiD: https://orcid.org/0000-0002-4772-3674*

[#2]*DSc, Professor, Department of Polymer Chemistry and Chemical Technology, Samarkand State University, University Blv. 15, Samarkand, Uzbekistan*
*ORCiD: https://orcid.org/0000-0001-5838-3743*

[1]hilola.xolmirzaeva@mail.ru, [2]f-normurot@samdu.uz

*Abstract -* In this article studied the influence of various factors on the synthesis of nano carbons from walnut shells, apricot kernels, methane, natural gas and propane-butane fractions, and also checked the textural and sorption characteristics of the obtained nanocarbon. The catalytic activity of a catalyst containing $(CuO)_x*(CoO)_y*(NiO)_z*(Fe_2O_3)_k*(MoO_3)_m/HSZ$ prepared based on "sol-gel" technology for the implementation of processes was studied under differential reactor conditions.

The morphological composition of the samples was performed by scanning electron microscopy on an LEO EVO 50 HVP instrument equipped with a part of energy dispersive X-ray microanalysis. Energy-dispersive X-ray spectra were recorded at a working distance of 10 mm at 20 kV. The microstructure of the samples was investigated by the method of a glowing electron microscope on a Jem-2000 CX device. Using the BET method, the specific surface area of the carbon nanotubes was measured.

Statistical and morphological characteristics of material porosity were calculated. The specific surface area is 168.7 $m^2/g$, the specific pore volume is 0.456 $cm^3/g$, and the average hole diameter is 2.58 nm. In the case of walnut shells, this absorption band was observed at 1376 $cm^{-1}$ and 1064 $cm^{-1}$. When examining samples based on apricot kernels in the infrared spectrum at various temperatures in an inert nitrogen atmosphere, it can be seen that cellulose and lignin are characterized by a strong absorption band, and the treatment decreases with increasing temperature.

The infrared spectra of carbon nanotubes have the same half-width, which indicates a structural change in change in the infrared spectrum in the 1300-1600 $cm^{-1}$ region. Changes in the infrared spectrum in the bands of ≈1330 $cm^{-1}$ and ≈1590 $cm^{-1}$ indicate the presence of graphite in the composition. This suggests that multilayer pipes differ sharply from single-layer pipes. When analysing the infrared spectra of carbon, the change in the frequency of 1730 $cm^{-1}$ refers to the longitudinal vibration C=O, which indicates the presence of oxygen groups in the compound. At the same time, the influence of various factors on the rate of formation of nanocarbon from fractions of methane, natural gas and propane-butane was studied, and the optimal conditions for the process were proposed.

**Keywords —** *Walnut Shells, Apricot Kernels, Methane, Natural Gas, Propane-Butane Fractions, Texture, Sorption, Catalyst, Nanocarbon, Synthesis.*

## I. INTRODUCTION

One of the most important inventions of the twentieth century on earth is the graphene or fullerene hydrocarbon nanotube. Until now, carbon was considered to have only two crystalline lattice graphitizations. Current observations have shown that carbon also has multilayer structures and that they can be pentagonal, hexagonal, hexagonal, or octagonal.

In the late twentieth century, new forms of fullerene and graphene nanotubes appeared. One of the most promising areas of nanotechnology is the synthesis of carbon nanomaterials. Multilayer carbon nanotubes occupy a special place among carbon nanomaterials. They are 10 to 80 nm in diameter and up to a few microns long. However, the addition of nanocarbon to materials leads to drastic changes in their physical and chemical properties. For example, the addition of 10% nanocarbon to polyethene leads to a strengthening of the material, changes in its hardness and melting temperature, as well as changes in the temperature range of transition to a highly elastic state [1-4].

To date, activated charcoal is used in many processes in the chemical industry, in the purification of wastewater and gas, in the production of alcohol products and in the medical industry. Basically, it is possible to achieve the development of adsorption technology by improving the method of production, which contributes to the constant improvement of the quality of the product. Various plant materials can be used as raw materials for the production of activated carbon: feldspar, peat, various types of walnut shells, fruit seeds and others. Also, carbonated materials are coal, hard coal, peat coke [5, 6]. It should be noted that the cost of imported HSZ for drying and saturation of natural gas is 1.5-2 times higher than domestic analogues. It is important to study the physicochemical and adsorption properties of adsorbents based on local raw materials instead of imported silica gels. Low-temperature gas separation technology is discussed in detail [7-14].

The choice of a reasonable method of condensate separation depends on the separation pressure and temperature, the composition of the gas and the conditions of its transport [15-17].

The problem of drying natural gas with various adsorbents has been discussed in several publications, in particular in the works [18-21].

At present, high-silicon zeolites are used in the physical and chemical processing of bentonite from Navbahor district of Uzbekistan, purification and drying of natural gas, oil and as a catalyst and catalyst holding material in the synthesis of many basic organic and petrochemicals [22-28].

The purpose of this work is to study the process of obtaining nano carbonated from walnut, apricot peel, natural gas and propane-butane fractions, as well as to study the texture and sorption properties of the obtained nanoglycerol.

## II. MATERIALS AND METHODS

To obtain nanocarbon from natural gas, we placed a $(CuO)_x*(CoO)_y*(NiO)_z*(Fe_2O_3)_k*(MoO_3)_m/HSZ$ based catalyst based on local raw materials in a 25 mm diameter quartz reactor, heated it to 650 °C in a 300 ml/min nitrogen stream and carried methane at 300 ml/min for 5 h. After synthesizing the product, we cooled it to 400 °C in nitrogen flow (300 ml/min) and room temperature in atmospheric air. A laboratory device for obtaining nanocarbon from methane is shown (Figure 1).

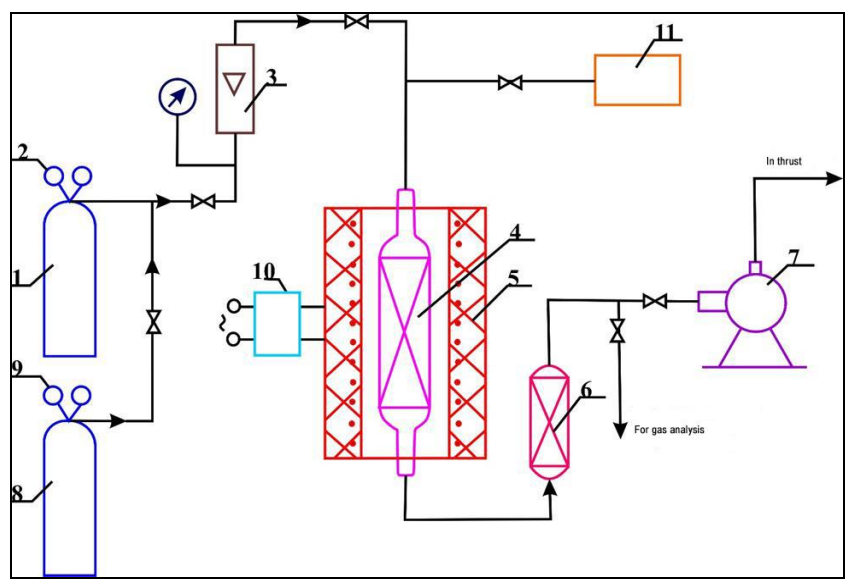

*1 natural gas. 2.9- reducer. 3 electronic gas flow meters. 4-reactor. 5 electric heater. 6- gas clock. 7 gas hours. 8- nitrogen balloon. 10- laboratory autotransformer (LATER). 11 Hydrogen production system*

**Fig. 1 Device for obtaining carbon nanotubes from natural gas in the laboratory**

The catalyst was prepared by the Sol-gel method as follows: to the dry mixture containing 20 g. $Cu(NO_3)_2·6H_2O$, 7.40 g. $Co(NO_3)_2·6H_2O$, 8.4 g. $Ni(NO_3)_2·6H_2O$, 9.6 g $Fe(NO_3)_2·9H_2O$, 13.68g $(NH_4)_2Mo_2O_7$ was added 60 ml of water and a mixture containing 33.20 g of glycerin and 45.36 g of citric acid. The mixture was stirred vigorously to form a viscous black-raspberry-coloured mass, which was placed in a porcelain bowl and treated at 250 °C -260 °C for 5 h. The result is a white-grey powder. To remove organic residue, we heated it at 500 °C for an additional 5 hours. We placed the obtained catalyst in a tubular quartz reactor with a diameter of 150 mm and carried out nanocarbon synthesis. To remove catalytic compounds, nanocarbon was boiled in concentrated HCl for 3 h, filtered, and washed with distilled water until neutral rN = 7. For filtration and washing, we used a Teflon filter funnel. The result is a catalyst containing $(CuO)_x*(CoO)_y*(NiO)_z*(Fe_2O_3)_k*(MoO_3)_m/HSZ$. The experiment used natural gas and propane-butane fraction, which are the main components of oil gases.

In the experiments, the catalyst containing $(CuO)_x*(CoO)_y*(NiO)_z*(Fe_2O_3)_k*(MoO_3)_m/HSZ$ was very active, and its decomposition into nanoparticles of methane, hydrogen, and carbon showed sufficient stability in reaction. The catalytic properties of a catalyst containing $(CuO)_x*(CoO)_y*(NiO)_z*(Fe_2O_3)_k*(MoO_3)_m/HSZ$ were investigated for the decomposition of methane at various thermal levels in a flow reactor.

The morphological composition of the samples was performed by the SEM method on an LEO EVO 50 XVP instrument equipped with an energy-dispersion X-ray microanalysis unit. Microphotography recording was performed at an operating distance of 5–25 kV and 8–10 mm under accelerating voltage. EDX spectra were recorded at 20 kV at a working distance of 10 mm. The microstructure of the samples was examined using the illuminated electron microscope on the instrument "JEM-2000 CX". The device "JEM-2000 CX" used an accelerated voltage of 200 kV. The catalysts were preheated and inactivated at 400 °C for 30 h at a flow rate of 30 ml/min of nitrogen. The processed catalysts were also passivated at room temperature after the reaction in the oxygen-argon mixture had stopped. We calculated the average size of 500 particles for each catalyst. A carbon nanotube was processed to measure the average diameter of a carbon nanotube. In addition to carbon nanotubes, graphite and metal catalyst particles, mainly in the form of amorphous carbon, may be present in the carbon material formed by the purification and separation of carbon nanotubes from the carbon material.

Studies have been conducted on the cleaning and recycling of carbon nanotubes to remove these contaminants. Various chemical and physicochemical methods of cleaning and separating nanotubes are now widely used in the literature. Certain methods are not without their drawbacks, the main ones being the complexity of the hardware design, the high cost, and the inability to process large amounts of carbon material.

According to the traditional method of production of activated charcoal, it consists of the process of activation by means of carbonizing and activating agents (inert gases, carbon dioxide, air, water vapour, etc.). This process is carried out by chemical and physical (vapour-gas) methods. Chemical methods include methods of handling raw materials with inorganic activators ($ZnCl_2$, $H_3PO_4$, HCl, $H_2SO_4$, etc.). The carbonization process of the samples was carried out under isothermal conditions. Heat treatment of the raw material was carried out in an inert gas atmosphere in the range of 650-700 °C, heating rate 15-20 °C/min and retention time 60 minutes (at a certain temperature). In the next step, to form micro-pores in the internal structure of the coal and thus increase the specific surface area, the obtained coal was steam-activated in an activation device at 800-850 60 for 60 minutes. Gas consumption was 1 litre per 200 g of adsorbent.

## II. RESULTS AND DISCUSSION

A photo of the charcoal taken from the walnut shell is shown in the figure below (Fig. 2).

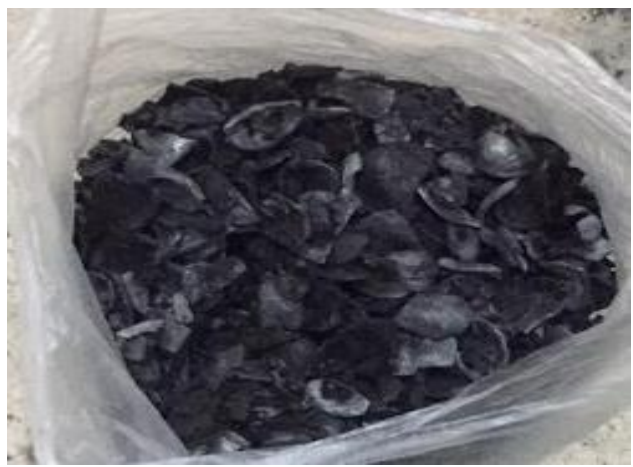

**Fig. 2 Coal from walnut shells**

In catalysts based on carbon nanotubes, after synthesis and functionalization of the carriers, the carbon nanotubes are condensed; their outer diameter remained unchanged and was 10–30 nm (Fig. 3). The diffractogram of a given carbon nanotube (Figure 4) shows one of the clearest and narrowest peaks in the 26° area, defined as the graphite peak. It should be noted that there is no asymmetric peak in the range $2\theta = 11.7°$, which is characteristic of amorphous carbon structures; Low-density peaks at angles of 44° and 54° may belong to multilayer carbon nanotubes. The specific surface area of carbon nanotubes was measured by the BET method from physical adsorption isotherms

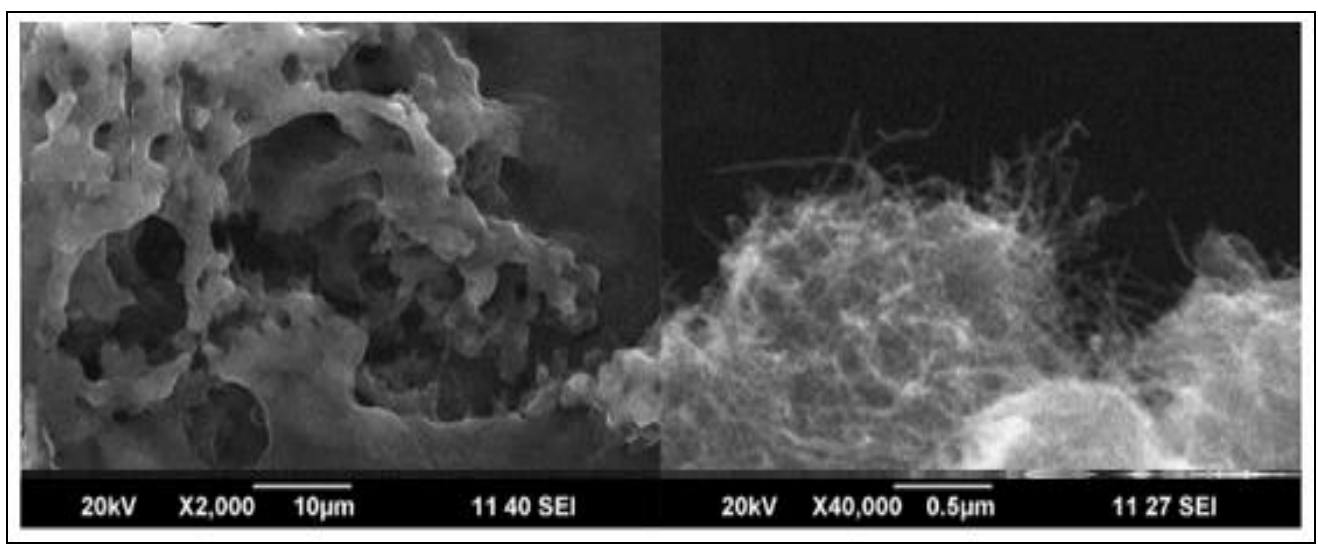

**Fig. 3 SEM Microphotography of Carbon Nanotube**

Statistical morphological characteristics of the porosity of the material were calculated: The specific surface area of BET is 168.7 m$^2$/g, specific pore volume 0.456 cm$^3$/g, the average hole diameter is 2.58 nm.

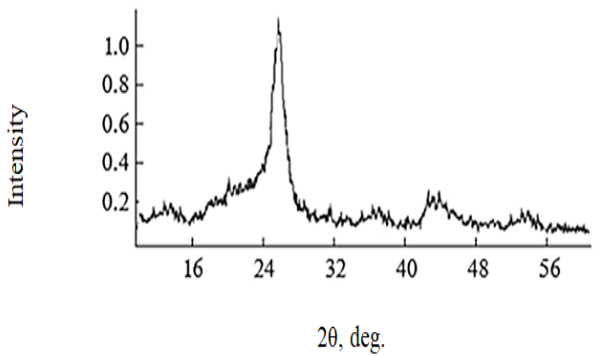

**Fig. 4 XRF-diffractogram of an impurity-free carbon nanotube**

Samples of carbon nanotubes synthesized to determine thermal properties were TGA scanned in a nitrogen atmosphere. It can be seen from the TGA results that weight loss is observed due to evaporation of moisture at temperatures below 100 °C and T = 250 °C.

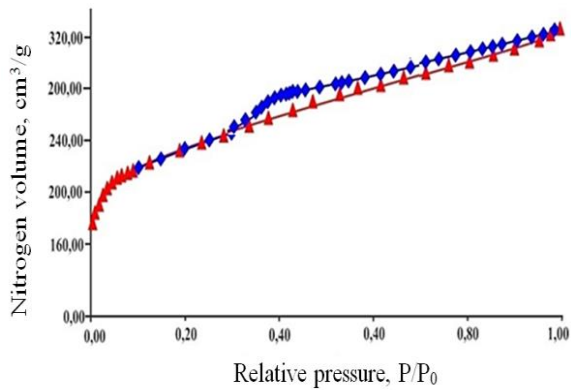

**Fig. 5 Activated carbon nitrogen adsorption isotherms derived from walnut shell**

Since carbon components are present in several forms in carbon nanotubes, the decomposition process proceeds in 2 stages: in the first stage (up to 825 °C), the more active phase (amorphous carbon) decomposes, and in the second (above 840 °C), carbon nanotubes decompose.

The residual mass at 998.4 °C is 73.35%. Examination of the structure of powdered materials by the low-temperature nitrogen adsorption method showed that samples obtained based on apricot peel have an improved micro powder structure. Activated carbon obtained from the walnut shell is in the form of miso powder, which is characterized by hysteresis in the adsorption isotherm (Fig. 5). It should be noted that the activation of carbons in the vapour-gas phase plays a key role in the formation of a powdery structure.

IRs were studied to determine the functional groups and structure of walnut and apricot peel. IRs determines the absorption band of the three hydroxyl groups in each glucopyranose network. In the case of walnut husks, this absorption band was observed in the 1376 cm$^{-1}$ and 1064 cm$^{-1}$ areas. When apricot peel-based samples were studied in IR at different temperatures of the inert nitrogen atmosphere, it was found that a strong absorption band is characteristic of cellulose and lignin, and processing decreases with increasing temperature (Fig. 6 a, b).

The presence of xylene in the sample, which is the main component of hemicellulose, showed that the absorption band 1737 cm$^{-1}$ belonged to the esters of the bitter uranium group, the intensity of which is characterized by the vibration of the C=O bond. If this fact is possibly explained, the process of carbonization reaction of condensed gas and solid pyrolysis of apricot peel leads to the formation of carbon-retaining microcrystals.

In these samples taken at high temperatures, the bands are completely absorbed. The observed changes may be related to the thermal diffusion (dissolution) of hemicellulose and its main component, xylan. The elemental composition of the samples was carried out by X-ray structural analysis. Radiation energy was 0-20 keV. The sensitivity of the method was 0.1 mass.%. The relative error did not exceed 10%.

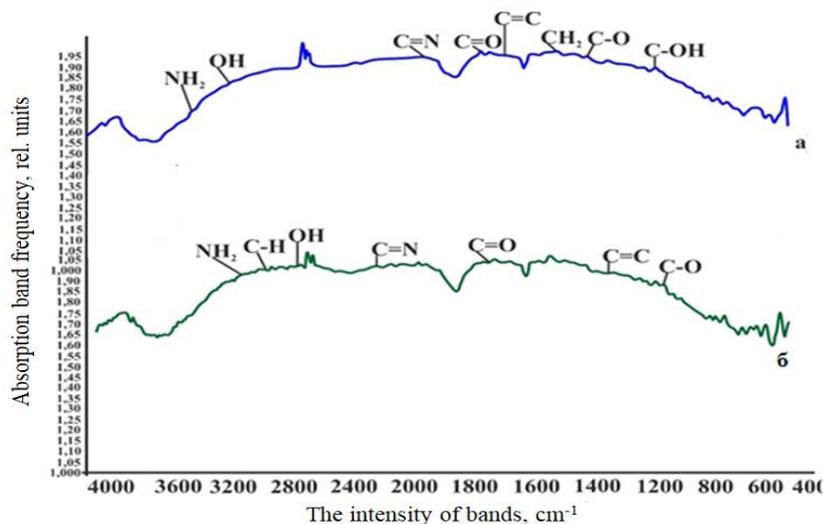

**Fig. 6. Infrared Spectra of Walnut And Apricot Peel**

In the experiment, the activity of carbon nanotubes was evaluated depending on the change in IR half-width and intensity. The half-widths of IRs of carbon nanotubes are the same, and the band intensities in the 1300–1600 cm$^{-1}$ region differ from each other. Since the change in intensity represents the oscillation of the structure, it indicates a structural change in the IR in the area of 1300–1600 cm$^{-1}$.

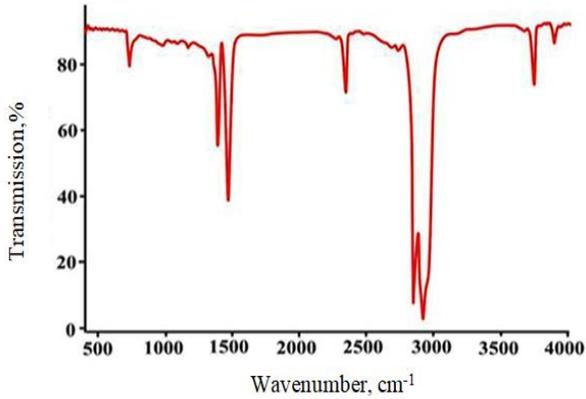

**Fig. 7 Infrared Spectra of Nanocarbon Synthesized From Walnut Shell**

Figures 7 and 8 show the presence of graphite in the composition of the changes in the bands IR i ≈1330 cm$^{-1}$ and ≈1590 cm$^{-1}$.

This indicates that multilayer pipes are drastically different from single-layer pipes. In the analysis of the infrared spectra of carbon, the change in frequency of 1730 cm$^{-1}$ is related to the longitudinal oscillation C = O, indicating the presence of oxygen groups in the composition of this mixture.

Figure 8 shows the electronic data on the fibrous structure of multilayer carbon tubes. This type of carbon nanotube is of the crystallization type with a fibre thickness of 0.03-0.1 μm and a length of several times the thickness of ≈5 μm.

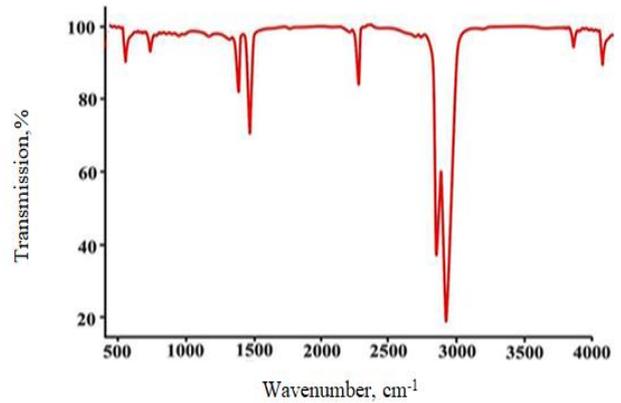

**Fig. 8 Infrared Spectra of Nanocarbon Synthesized From Apricot Peel**

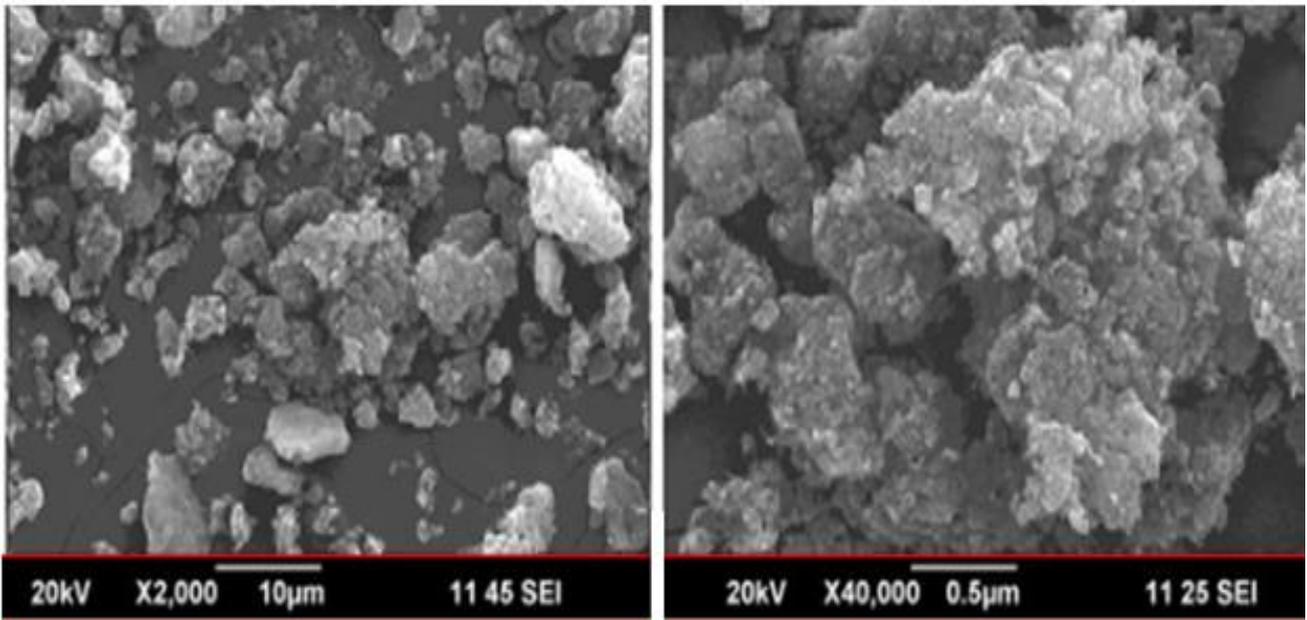

**Fig. 9. Electronic Data of The Fibrous Structure Of Multilayer Carbon Tubes**

Carbon nanotubes form aggregates of different sizes from 5 μm to 2.5 μm in the form of bundles. Aggregates are formed based on Van der Waals forces. It was observed that the multilayer nanotubes consisted of homogeneous and other modifications of amorphous carbon fullerene particles (Fig. 9).

Electronic images of homogeneous carbon tubes show that the morphological structure of a single-layer tube differs from that of a multilayer (Fig. 9). In this case, the carbon nanotube consists of spherical crystals with a size of ≈0.01 μm. The individual nanoparticles will be densely packed. The carbon nanotube forms particles from 2 μm to 120 μm due to Van der Waals forces. The studied carbon nanotube has a microstructure of different types of media: multilayer carbon tube - irregular, layered, fibrous, pseudo-spherical arrangement for single-layer carbon tubes.

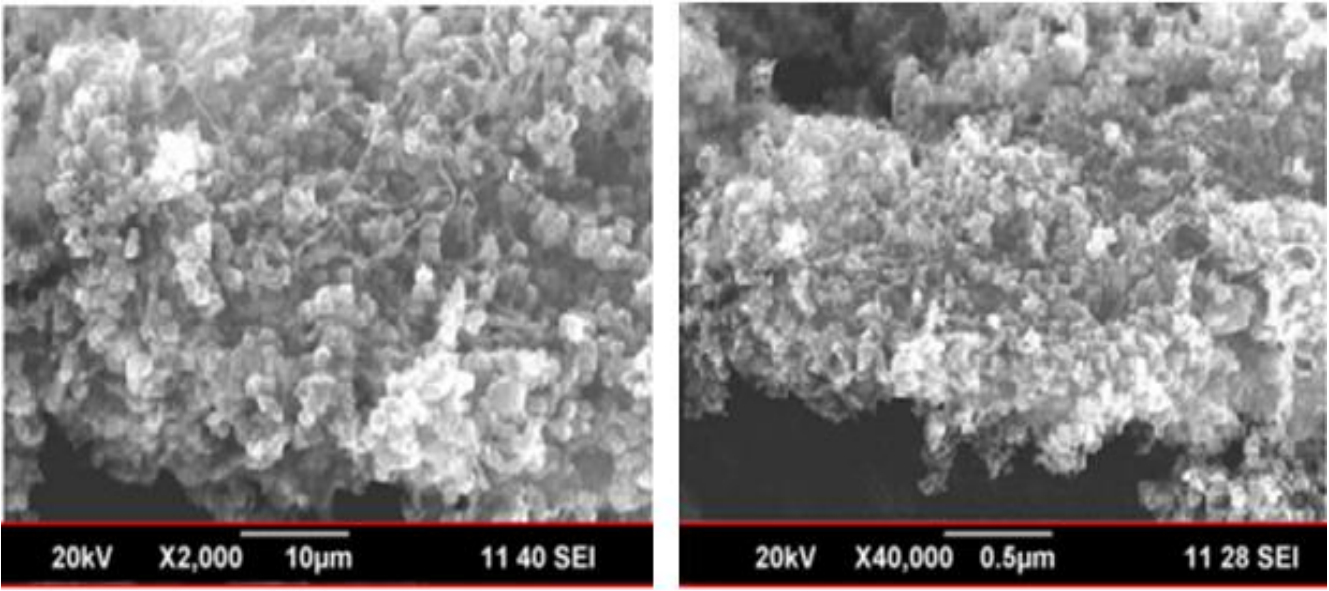

**Fig. 10 Electronic Images of Homogeneous Carbon Tubes**

Then conducted a series of experiments to obtain nano carbons from natural gas, methane, and propane-butane fractions in the presence of a catalyst containing $(CuO)_x*(CoO)_y*(NiO)_z*(Fe_2O_3)_k*(MoO_3)_m/HSZ$.

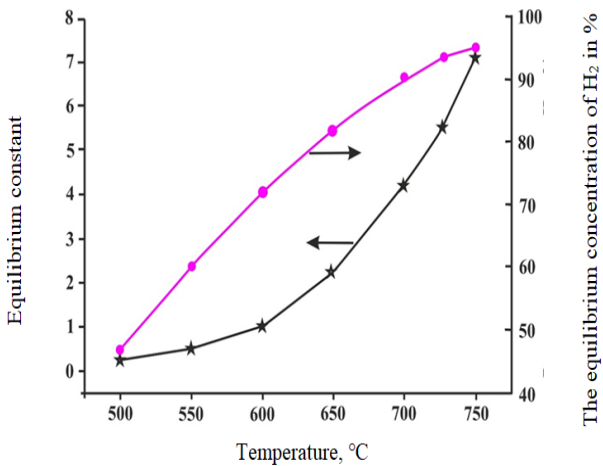

**Fig. 11 Temperature Dependence of The Decomposition of Methane Into Carbon and Hydrogen**

From the results of the experiment, it can be concluded that although the catalyst $(CuO)_x*(CoO)_y*(NiO)_z*(Fe_2O_3)_k*(MoO_3)_m/HSZ$ has high activity and stability, there are thermodynamic limits in the reactions that decompose methane. Figure 10 shows the temperature dependence of the decomposition of methane into carbon and hydrogen.

Kinetic calculations of the formation of carbon from methane $(CuO)_x*(CoO)_y*(NiO)_z*(Fe_2O_3)_k*(MoO_3)_m/HSZ$ at 700 °C -750 °C in the catalyst, as well as the growth of carbon nanoparticles in the catalyst, appears to be moderate (Fig. 11).

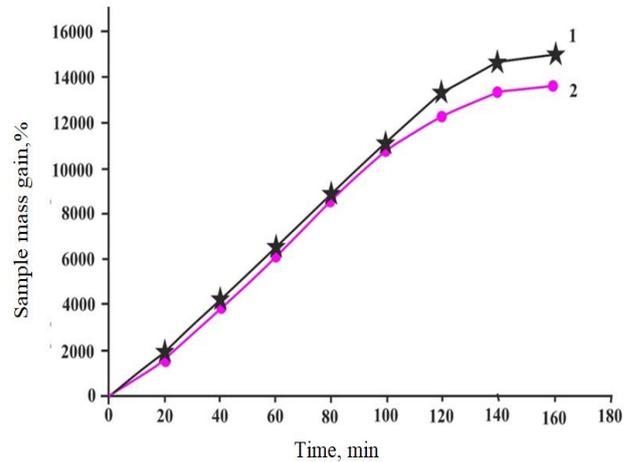

*The reaction temperature is 700 °C. Delivery of natural gas at -10 l/s. Weight of $(CuO)_x*(CoO)_y*(NiO)_z*(Fe_2O_3)_k*(MoO_3)_m/HSZ$ catalyst is 0.5 g.*

*The red dotted line indicates the equilibrium hydrogen concentration level.*

**Fig. 14. Time dependence of hydrogen concentration at reactor output (%).**

During the steady-state of the reaction, the volume of hydrogen released from the reactor was 70%. The red dots indicate the level of hydrogen. This means that the volume of hydrogen in the stationary period at the outlet of the reactor is about 25% lower (Figure 14).

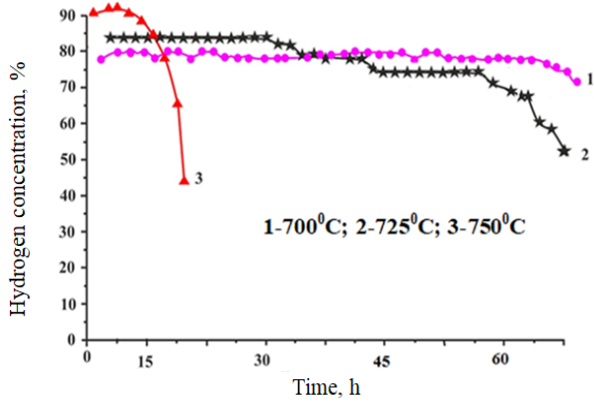

Delivery of natural gas at -10 l/s. The mass of the $(CuO)_x*(CoO)_y*(NiO)_z*(Fe_2O_3)_k*(MoO_3)_m$/HSZ catalyst is 5 g.

**Fig. 15 Variation of Hydrogen Concentration (%) with the Temperature at the Outlet of the Reactor**

As can be seen in Figure 15, the volume of hydrogen increased by the temperature at the exit of the reactor at 700 °C, while the volume of hydrogen in the stationary period of the reaction was 70% at the exit of the reactor at 725° increased by 75% at 750°, to 84% at 750°. But the regular operation of the catalyst led to a decrease in the rate of cancer without increasing. The stationary reaction period of 750 °C was 4-5 hours.

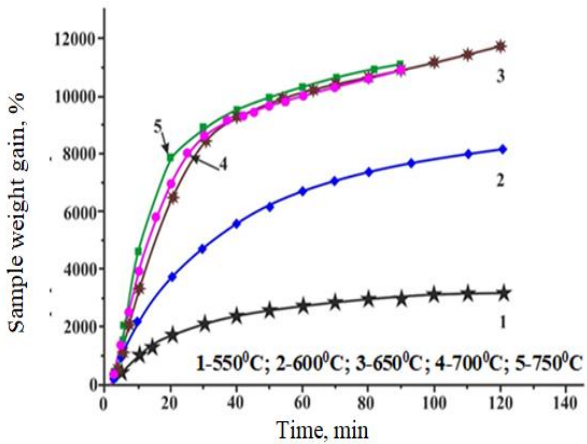

**Fig. 16 Kinetic Curve Lines of Carbon Release in $(CuO)_x*(CoO)_y*(NiO)_z*(Fe_2O_3)_k*(MoO_3)_m$/HSZ catalyst from C3H8-C4H10 Mixture In Different Temperatures**

The kinetic lines of carbon release from the propane-butane mixture at different levels under the action of the $(CuO)_x*(CoO)_y*(NiO)_z*(Fe_2O_3)_k*(MoO_3)_m$/HSZ catalyst are given (Fig. 15). The optimum level for the reaction is 650 °C -750 °C.

As the rate of the reaction increased, the carbon release remained unchanged at 110–120 g/g (Figure 16). In the front layer of the $(CuO)_x*(CoO)_y*(NiO)_z*(Fe_2O_3)_k*(MoO_3)_m$/HSZ catalyst, as it enters the reactor, the propane-butane mixture decomposes to form carbon nanoparticles and hydrogen. In the next zones of the reactor, there is a mixture of hydrogen and hydrocarbons, and we observed how hydrogen affects the formation of carbon. We diluted the $C_3H_8$-$C_4H_{10}$ mixture with hydrogen and found that the growth rate of the primary carbon nanoparticles decreased. But as a result, carbon emissions increased by 250-260 g/g. The pyrolysis process in the propane-butane mixture in the empty reactor is also very active, as can be seen from the reaction. Pyrolysis also produces olefins, ethylene, propylene and butylene. They can form more active nanoparticles than saturated hydrocarbons. As the reaction temperature rose to 700 °C, the decomposition of the propane-butane mixture increased. At the exit of the reactor, the volume of hydrogen was 71-72 mol%, and the volume of methane was 25-26 mol%. The normal operation of the catalyst is 13-14 hours. The result of the release of carbon nanoparticles after 24 hours is 320 g/g. When the reaction temperature rises to 750 °C, the volume of hydrogen leaving the reactor increases by 81-82 mol%, and the volume of methane decreases by 16-17 mol%. However, the normal operation of the $(CuO)_x*(CoO)_y*(NiO)_z*(Fe_2O_3)_k*(MoO_3)_m$/HSZ catalyst decreased by 5-5.5 hours.

The volume rate of the propane-butane mixture was given as 12,000 $h^{-1}$. After 8 hours of reaction, the carbon nanoparticle yield was 50 g/g. It follows that the optimum level for the formation of hydrogen and carbon nanoparticles from propane-butane compounds under the action of $(CuO)_x*(CoO)_y*(NiO)_z*(Fe_2O_3)_k*(MoO_3)_m$/HSZ catalyst is 700-725 °C. If this reaction takes place under optimal conditions, 700-750 °C litres of hydrogen and 320-350 g of carbon nanoparticles can be obtained from 1 g of $(CuO)_x*(CoO)_y*(NiO)_z*(Fe_2O_3)_k*(MoO_3)_m$/HSZ catalyst in a propane-butane mixture. The results of the chromatographic analysis of the gas composition formed when the propane-butane mixture reacted at 750 °C for 200 min (Figure 17).

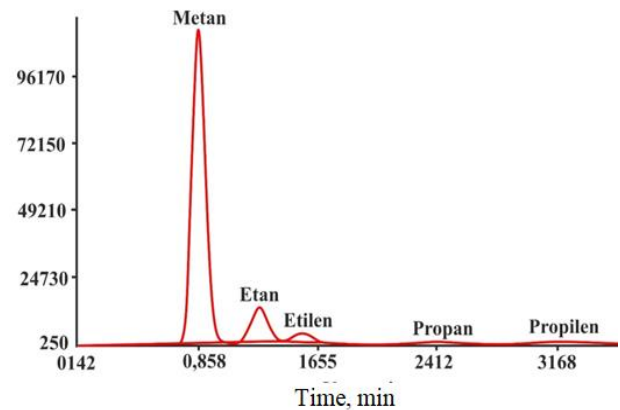

**Fig. 17 Chromatogram of hydrocarbon products formed during the decomposition of the mixture C3H8-C4H10: Treac.=750 °C (τreac=200 min) Catalyst-$(CuO)_x*(CoO)_y*(NiO)_z*(Fe_2O_3)_k*(MoO_3)_m$/HSZ**

An SEM microphotograph of the nanocarbon formed by decomposing the propane-butane mixture on a catalyst formed at 700 °C is shown (Fig. 18).

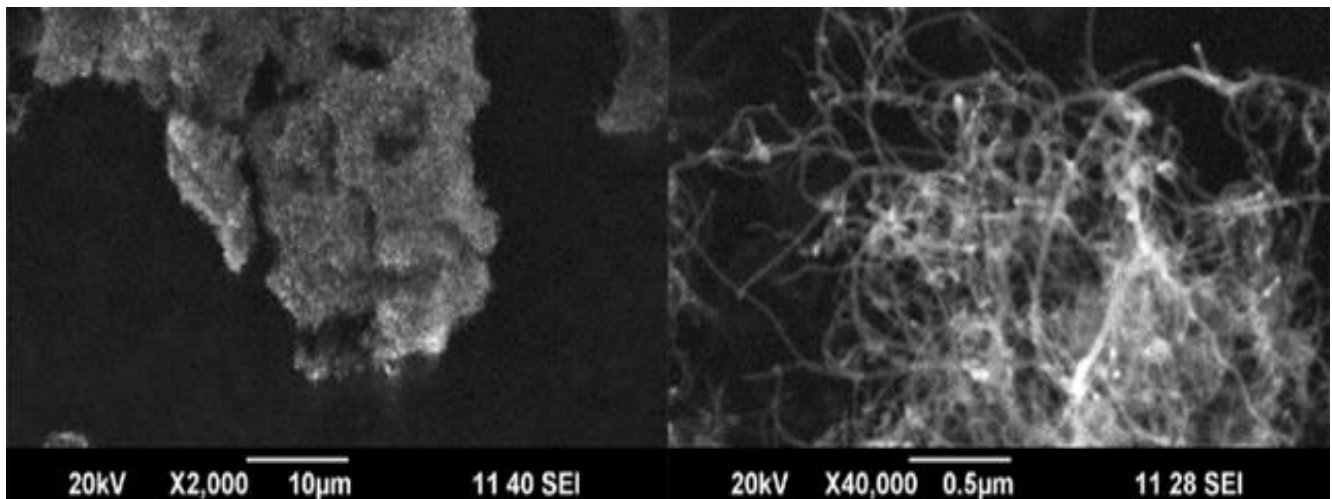

**Fig. 18 Scanning Electron Microscopy Microphotography of Nanocarbon Formed By the Decomposition of Propane-Butane Mixture on Catalyst $(CuO)_x*(CoO)_y*(NiO)_z*(Fe_2O_3)_k*(MoO_3)_m$/HSZ, 700 °C**

Determining the amount of carbon in a carbonized sample is an important task. To this end, we conducted a series of studies on the amount of carbon in the synthesized sorbents. To do this, we heat-treated the sorbent at 700–800 °C for two hours. It is known that the main indicator that characterizes the efficiency of sorbents is their sorption capacity. To determine the sorption capacity of the sorbents, we determined the absorption of oil and gasoline by the sorbents.

Adsorption was studied in a specially designed device consisting of a glass tube with a length of 25 cm and an inner diameter of 5 mm and 20 ml vials with a magnetic cap. The vial is equipped with inlet and outlet ports. The data obtained for the determination of the sorption capacity of sorbents are given in Table 1.

**Table 1. Sorption capacity of sorbents.**

| Heat treatment temperature of sorbents | Sorption time, hours | Sorption capacity, g/g | |
|---|---|---|---|
| | | On oil | On gasoline |
| 750 °C | 1 | 0,58 | 0,28 |
| | 2 | 0,64 | 0,24 |
| | 24 (a day) | 0,92 | 0,29 |
| 800 °C | 1 | 0,46 | 0,22 |
| | 2 | 0,48 | 0,28 |
| | 24 (a day) | 0,86 | 0,32 |

### III. CONCLUSION

1. The process of production of activated carbon was carried out in two stages: pyrolysis (carbonization) and activation (activators: inert gases, carbon dioxide, air, water vapour, etc.) by chemical and steam-gas methods. Temperature regimes have been optimized for the production of nanocarbon from fractions of walnut shell, apricot kernel, methane, natural gas and propane-butane.

2. The effect of various factors on the synthesis of nanocarbon from the fractions of walnut peel, apricot kernel, methane, natural gas and propane-butane was studied, as well as the texture and sorption characteristics of the obtained nanocarbon were examined.

3. The catalytic activity of a catalyst containing $(CuO)_x*(CoO)_y*(NiO)_z*(Fe_2O_3)_k*(MoO_3)_m$/HSZ, prepared based on "Sol-gel" technology for the implementation of processes, was studied under differential reactor conditions.

4. The morphological composition of nanocarbon was measured by SEM method, electron microscope method with illuminated microstructure, the specific surface area of nanotubes was measured by BET method from physical adsorption isotherms.

5. Statistical morphological characteristics of the porosity of the material were calculated: The specific surface area of BET is 168.7 $m^2/g$, specific pore volume 0.456 $cm^3/g$, the average hole diameter is 2.58 nm. IRs were observed in the nanocarbon absorption band 1376 $cm^{-1}$ and 1064 $cm^{-1}$ in the walnut shell. When apricot peel-based samples were studied at different temperatures in an inert nitrogen atmosphere in IR, it was observed that the strong absorption band is characteristic of cellulose and lignin and that processing decreases with increasing temperature. The half-widths of the IRs of carbon nanotubes are the same, indicating a structural change in the IR in the 1300-1600 $cm^{-1}$ region. The changes in the IR ≈1330 $cm^{-1}$ and ≈1590 $cm^{-1}$ bands indicate the presence of graphite.

6. The effect of various factors on the rate of formation of nanocarbon obtained from methane, natural gas and propane-butane fractions was studied, and optimal process conditions were proposed.

### ACKNOWLEDGEMENT
We take this opportunity to thank all the people who have supported and guided us during the completion of this work.

Conflict of Interest**:** The authors report no conflicts of interest.

Financing: Source of funding is nil.